\documentclass[a4paper,12pt]{article}

\usepackage{epsfig}
\usepackage{amssymb,amsmath}
\newcommand{\tr}{\,\mbox{Tr}\,}
\newcommand{\lapprox}{%
\mathrel{%
\setbox0=\hbox{$<$}
\raise0.6ex\copy0\kern-\wd0
\lower0.65ex\hbox{$\sim$}
}}
\newcommand{\gapprox}{%
\mathrel{%
\setbox0=\hbox{$>$}
\raise0.6ex\copy0\kern-\wd0
\lower0.65ex\hbox{$\sim$}
}}

\newcommand{\ba}{\begin{array}}
\newcommand{\ea}{\end{array}}
\newcommand{\bd}{\begin{displaymath}}
\newcommand{\ed}{\end{displaymath}}
\newcommand{\be}{\begin{equation}}
\newcommand{\ee}{\end{equation}}
\newcommand{\bea}{\begin{eqnarray}}
\newcommand{\eea}{\end{eqnarray}}

\newcommand{\mm}{\mathcal{M}}
\newcommand{\diag}{\mbox{diag}\,}
\def\q2 {q^2}

\def\bt{\begin{table}}
\def\et{\end{table}}

\catcode`@=11 % This allows us to modify PLAIN macros.
\def \gsim{\mathrel{\mathpalette\@versim>}}
\def \lsim{\mathrel{\mathpalette\@versim<}}
\def \@versim#1#2{\lower0.4ex\vbox{\baselineskip\z@skip\lineskip\z@skip
     \lineskiplimit\z@\ialign{$\m@th#1\hfil##\hfil$%
     \crcr#2\crcr\sim\crcr}}}
\catcode`@=12 % at signs are no longer letters

\begin{document}

\begin{flushright}
{\small 
RECAPP-HRI-2013-005 \\
UWThPh-2013-10}
\end{flushright}

\begin{center}

{\large\bf Doubly charged scalar decays in a type~II seesaw scenario with two
  Higgs triplets}\\[15mm] 
Avinanda Chaudhuri $^{a,}$\footnote{E-mail: avinanda@hri.res.in},
Walter Grimus $^{b,}$\footnote{E-mail: walter.grimus@univie.ac.at}
and Biswarup Mukhopadhyaya $^{a,}$\footnote{E-mail: biswarup@hri.res.in} 
\\[2mm]
{\em $^a$Regional Centre for Accelerator-based Particle Physics \\
     Harish-Chandra Research Institute
\\
Chhatnag Road, Jhusi, Allahabad - 211 019, India}
\\[2mm]
{\em $^b$University of Vienna, Faculty of Physics\\
     Boltzmanngasse 5, 1090 Vienna, Austria }\\
[20mm] 
\end{center}

\begin{abstract} 
The type~II seesaw mechanism for neutrino mass generation usually
makes use of one complex scalar triplet. The collider signature
of the doubly-charged scalar, the most striking feature of this scenario,
consists mostly in decays into same-sign dileptons or same-sign $W$~boson pairs.
However, certain scenarios of neutrino mass generation, such as
those imposing texture zeros by a symmetry mechanism, 
require at least two triplets
in order to be consistent with 
the type~II seesaw mechanism. 
We develop a model with two such complex triplets and show that, in such a case,
mixing between the triplets can cause the heavier doubly-charged 
scalar mass eigenstate to decay into a singly-charged scalar and a
$W$~boson of the same sign. Considering a large number of benchmark points 
with different orders of magnitude of the $\Delta L =2$ Yukawa couplings,
chosen in agreement with the observed 
neutrino mass and mixing pattern, we 
demonstrate that $H^{++}_1 \rightarrow H^+_2 W^+$ can have more than
99\% branching fraction 
in the cases where the vacuum expectation values of the triplets 
are small. 
It is also shown that the above decay allows one to differentiate a
two-triplet case 
at the LHC, through the ratios of events in various
multi-lepton channels. 
\end{abstract}

\vskip 1 true cm

\newpage
\setcounter{footnote}{0}

\def\baselinestretch{1.5}
%==============================================================================
%==============================================================================
\section{Introduction}
It is by and large agreed that the Large Hadron Collider (LHC)
has discovered the Higgs boson predicted in the standard electroweak
theory, or at any rate a particle with close resemblance to 
it~\cite{ATLAS1,CMS1}. 
At the same time, driven by both curiosity and various physics
motivations, physicists have been exploring the possibility that the
scalar sector of elementary particles contains more members than just
a single $SU(2)$ doublet. A rather well-motivated scenario often
discussed in this context is one containing at least one complex
scalar $SU(2)$ triplet of the type 
$(\Delta^{++}, \Delta^{+}, \Delta^{0})$~\cite{kummer}.  
A small vacuum expectation value of the neutral member
of the triplet, constrained as it is by the $\rho$-parameter, can lead
to Majorana masses for neutrinos, driven by $\Delta L = 2$ Yukawa
interactions of the triplet. Such mass generation does not require any
right-handed neutrino, and this is the quintessential principle of the
type~II seesaw mechanism~\cite{II,generalD+M}.

One of the most phenomenologically striking features of this mechanism is the
occurrence of a doubly-charged scalar. Its signature at TeV scale
colliders is expected to be seen, if the triplet masses are not too
far above the electroweak symmetry breaking scale. The most conspicuous
signal consists
in the decay into a pair of same-sign leptons, 
i.e. $\Delta^{++} \rightarrow \ell^+ \ell^+$. 
The same-sign dilepton invariant mass peaks resulting from this
make the doubly-charged scalar show up rather
conspicuously. Alternatively, the decay
into a pair of same-sign $W$ bosons, i.e. $\Delta^{++} \rightarrow W^+ W^+$,  
is  dominant in a
complementary region of the parameter space, which---though more
challenging from the viewpoint of background elimination---can unravel
a doubly-charged scalar~\cite{PD}.

In this paper, we shall discuss the
situation where a third decay channel, namely a doubly-charged scalar
decaying into a singly-charged scalar and a $W$ of the same sign, is dominant
or substantial. Such a decay mode is usually suppressed, since the
underlying $SU(2)$ invariance implies relatively small mass splitting
among the members of a triplet. However, when several triplets
of a similar nature are present and mixing among them is allowed, a
transition of the above kind is possible between two scalar mass
eigenstates. Apart from being interesting in itself, several scalar triplets
naturally occur in models for neutrino masses and lepton mixing based on the
type~II seesaw mechanism. In particular, it has been 
shown 
that in such a scenario a realization of viable neutrino mass
matrices with 
two texture zeros~\cite{frampton},\footnote{Texture zeros are a favorite means
  of achieving relations between masses and mixing angles, see for
  instance~\cite{tz}.} 
using symmetry arguments~\cite{GJLT}, requires 
two or three scalar triplets~\cite{grimus2}. 
In this paper, we take up the case of two coexisting triplets. 
We demonstrate
that in such cases one doubly-charged state can often decay into a
singly-charged state and a $W$ of identical charge.
This is not surprising, because each of the two erstwhile studied decay modes 
is controlled by parameters that are rather suppressed. In the case
of  $\Delta^{++} \rightarrow \ell^+ \ell^+$,  the amplitude is
proportional to the $\Delta L =2$ Yukawa coupling, while for
$\Delta^{++} \rightarrow W^+ W^+$,  it is driven by the
triplet vacuum expectation value (VEV).  The restrictions from 
neutrino masses as well as precision electroweak constraints
makes both of these rates rather small. On the other hand,
in the scenario with two scalar triplets with charged mass eigenstates
$H^{++}_k$ and $H^+_l$ ($k,l=1,2$),
the decay amplitude for $H_1^{++} \rightarrow H_2^+ W^+$,  
if kinematically allowed, is controlled by the $SU(2)$ gauge coupling.
Therefore, if one identifies regions of the parameter space where
it dominates, one needs to devise new search strategies at the 
LHC~\cite{huitu}, including ways of eliminating backgrounds.

We note that the mass parameters of the two triplets, on which no
phenomenological restrictions exist, are {\it a priori} unrelated and, 
therefore, as a result of mixing between the two triplets, 
the heavier doubly-charged state can decay into a lighter, singly-charged state
and a real $W$ over a wide range of the parameter space. In that range it is
expected that this decay channel dominates
for the heavier doubly-charged state.
By choosing a number of benchmark points, we demonstrate that this is
indeed the case. 

In section~2, we present a summary of the model with a single
triplet and explain why the decay 
$\Delta^{++} \rightarrow \Delta^{+} W^{+}$ is disfavoured there.  
The details of a two-triplet scenario,
including the 
scalar potential 
and the composition of the
physical states, are presented in section~\ref{2-triplet}. We select several
benchmark points and show the decay patterns of the corresponding
doubly-charged scalars in section~\ref{benchmark}, where their production
rates at the LHC are also presented. 
We point out the usefulness of $H_1^{++} \rightarrow H_2^+ W^+$
at the LHC in the context of our model with two scalar triplets in
section~\ref{usefulness}. 
We summarise and conclude in section~\ref{concl}.
In appendix~\ref{bps} the input parameters for the benchmark points
are listed while appendix~\ref{expressions} 
contains the formulas for the decay rates of the
doubly-charged scalars.

\section{The scenario with a single triplet}
\label{1-triplet}
In this section we perform a quick recapitulation of the scenario
with a single triplet field, in addition to the usual Higgs
doublet $\phi$, using the notation of~\cite{grimus3}.  The Higgs triplet 
$\Delta = (\Delta^{++}, \Delta^{+}, \Delta^0)$ is represented by
the $2 \times 2$ matrix
\begin{equation}\label{triplet}
\Delta = 
\left(\begin{array}{cc}
\Delta^+ & \sqrt{2}\Delta^{++} \\ \sqrt{2}\Delta^0 & -\Delta^+ 
\end{array}\right).
\end{equation}
The VEVs of the doublet and the triplet
are given by
\begin{equation}\label{vev}
\langle \phi \rangle_0 =
\frac{1}{\sqrt{2}} \left(\begin{array}{c} 0 \\ v \end{array}\right)
\quad \mbox{and} \quad
\langle \Delta \rangle_0 =
\left(\begin{array}{cc} 0 & 0 \\ w & 0 \end{array}\right),
\end{equation}
respectively. Thus, the triplet VEV is obtained as
$\langle \Delta^0 \rangle = w/\sqrt{2}$ . 
The only doublet-dominated physical
state that survives after the generation of gauge boson masses is a 
neutral scalar $H$.

The most general scalar potential involving $\phi$ and $\Delta$ can be
written as 
\begin{eqnarray}
V(\phi,\Delta)\  = \hphantom{+} a\, \phi^\dagger\phi + \ \frac{b}{2}
\tr(\Delta^\dagger\Delta)+ \ c\, (\phi^\dagger\phi)^2   
+ \ \frac{d}{4}\left( \tr(\Delta^\dagger\Delta) \right)^2 \nonumber \\ +
\ \frac{e - h}{2} \phi^\dagger\phi \tr(\Delta^\dagger\Delta)  
+  \ \frac{f}{4} \tr(\Delta^\dagger\Delta^\dagger)\tr(\Delta\Delta)
\nonumber \\ 
+ \ h \phi^\dagger \Delta^\dagger \Delta \phi + \left( t\, \phi^\dagger
\Delta \tilde{\phi} + 
\mbox{H.c.} \right),
\end{eqnarray}
where $\tilde{\phi} \equiv i\tau_2 \phi^\ast$. 
For simplicity, we assume both $v$ and $w$ to be 
real and positive, which requires $t$ to be real as well.
In other words, all CP-violating effects are neglected in this study.

The choice $a<0$, $b>0$ ensures that the primary source of spontaneous
symmetry breaking resides in the VEV of the scalar doublet.
Without any loss of generality, we assume the following orders of
magnitude for the parameters in the potential:
\begin{equation}\label{oomagn}
a,\: b \sim v^2; \quad
c,\: d,\: e,\: f,\: h \sim 1; \quad 
|t|\ll v.  
\end{equation}
Such a choice is motivated by 
\begin{enumerate}
\renewcommand{\labelenumi}{(\alph{enumi})}
\item
proper fulfillment of the
electroweak symmetry breaking conditions, 
\item
the need to have $w \ll v$ small due to
the $\rho$-parameter constraint, 
\item
the need to keep doublet-triplet mixing low in general,
and 
\item 
the urge to ensure perturbativity of all quartic couplings.  
\end{enumerate}

The mass Lagrangian for the singly-charged scalars in this model is given by
\begin{equation}
\mathcal{L}^\pm_S = -
(H^-, \phi^-) \mathcal{M}_+^2 
\left( \begin{array}{c} H^+ \\ \phi^+ \end{array} \right)
\end{equation}
with\footnote{Note that the matrix $\mathcal{M}_+^2$ given here is correct,
  whereas in equation~(42) of
  reference~\cite{grimus3} the 11 and 22-elements of the same mass matrix
  are exchanged by error.}
\begin{equation}
\mathcal{M}_+^2 = \left(\begin{array}{cc}
(q+h/2)v^2 & \sqrt{2} v (t - w h/2) \\
\sqrt{2} v (t - w h/2) & 2(q + h/2)w^2 
\end{array}\right) 
\quad \mbox{and} \quad 
q = \frac{|t|}{w}.
\end{equation}
The field $\phi^+$ is the charged component of the doublet scalar field of the
Standard Model (SM).
One of the eigenvalues of this matrix is zero corresponding to the
Goldstone boson which gives mass to the $W$ boson. The mass-squared of
the singly-charged physical scalar is obtained as
\begin{equation}
m^2_{\Delta^+} = \left(q + \frac{h}{2} \right) (v^2 + 2w^2),
\end{equation} 
whereas the corresponding expression for the doubly-charged scalar is
\begin{equation}
m^2_{\Delta^{++}} = (h + q) v^2 + 2fw^2.
\end{equation}
Thus, in the limit  $w \ll v$,
we obtain 
\begin{equation}
m^2_{\Delta^{++}} - m^2_{\Delta^+} \simeq {\frac {h}{2}}v^2.
\end{equation}
 
\begin{figure}%[h!t]
\centerline{
        \includegraphics[width=0.75\textwidth]{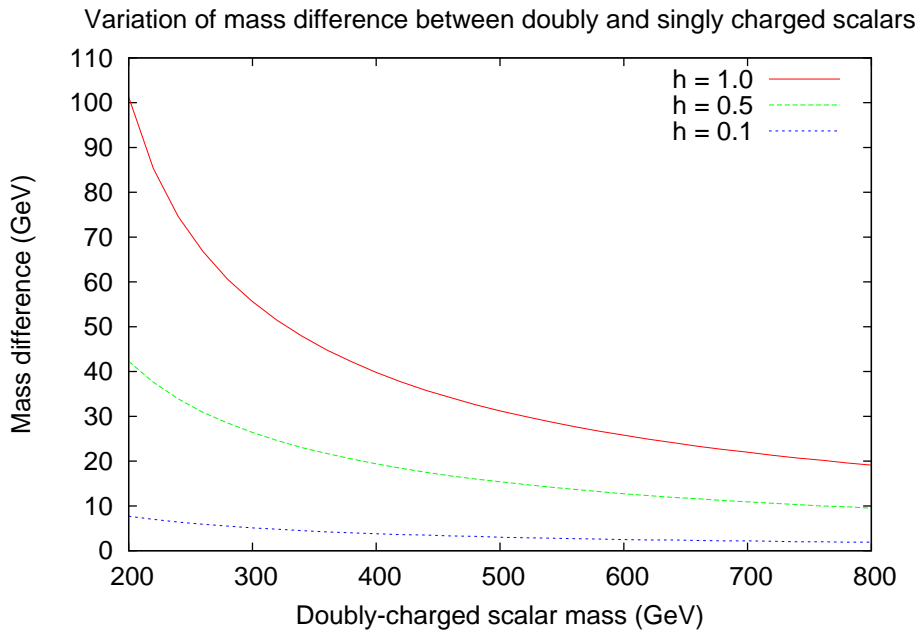}
}
\caption{Variation of mass difference between the doubly and singly-charged
  scalars, for various values of the parameter $h$.} 
\label{fig}
\end{figure}

It is obvious from the above that a substantial mass splitting 
between $\Delta^{++}$ and $\Delta^+$ is in general difficult.
This is clear from Figure~\ref{fig} where we plot the mass difference
between the two states for different values of $h$. Sufficient splitting,
so as to enable the decay  $\Delta^{++} \rightarrow \Delta^+ W^+$
to take place with appreciable branching ratio, will require
$h\simeq 1$,  $m_{\Delta^{++}} \lesssim 250$\,GeV and a correspondingly
smaller $m_{\Delta^+}$.  The limits from LEP and Tevatron 
disfavour triplet states with such low masses. Thus one concludes
that the phenomenon of the doubly-charged scalar decaying into
a singly-charged one and a $W$ is very unlikely.

\section{A two Higgs triplet scenario}
\label{2-triplet}
There may, however, be some situations where a single triplet
is phenomenologically inadequate. This happens, for example,
when one tries to impose texture zeros in the neutrino mass matrix
within a type~II seesaw framework by using Abelian
symmetries~\cite{grimus2}. Having this is in mind, 
we venture into a model consisting of  
one complex doublet and two 
$Y=2$ triplet scalars $\Delta_1$, $\Delta_2$, both written as 
2$\times$2 matrices:
\begin{equation}\label{triplet2}
\Delta_1 = 
\left(\begin{array}{cc}
\delta_1^+ & \sqrt{2}\delta_1^{++} \\ \sqrt{2}\delta_1^0 & -\delta_1^+ 
\end{array}\right)
\quad \mbox{and} \quad
\Delta_2 = 
\left(\begin{array}{cc}
\delta_2^+ & \sqrt{2}\delta_2^{++} \\ \sqrt{2}\delta_2^0 & -\delta_2^+ 
\end{array}\right).
\end{equation}
The VEVs of the scalar triplets are given by
\begin{equation}\label{vev2'}
\langle \Delta_1 \rangle_0 =
\left(\begin{array}{cc} 0 & 0 \\ w_1 & 0 \end{array}\right) 
\quad \mbox{and} \quad
\langle \Delta_2 \rangle_0 =
\left(\begin{array}{cc} 0 & 0 \\ w_2 & 0 \end{array}\right).
\end{equation}
The VEV of the Higgs doublet is as usual given by equation~(\ref{vev}).

The scalar potential in this model involving $\phi$, $\Delta_1$
and $\Delta_2$ can be written as 
\begin{eqnarray}
\lefteqn{V(\phi,\Delta_1,\Delta_2) =}
\nonumber \\
&&
a\, \phi^\dagger\phi + 
\frac{1}{2}\, b_{kl} \tr(\Delta_k^\dagger \Delta_l)+ 
c (\phi^\dagger\phi)^2 + 
\frac{1}{4}\, d_{kl} \left( \tr(\Delta_k^\dagger\Delta_l) \right)^2
\nonumber \\
&&
+ \frac{1}{2}\,(e_{kl} - h_{kl})\,
\phi^\dagger \phi  \tr (\Delta_k^\dagger\Delta_l) +
\frac{1}{4}\,f_{kl}  
\tr(\Delta_k^\dagger\Delta_l^\dagger) \tr(\Delta_k\Delta_l)
\nonumber \\
&&
+ h_{kl}\, \phi^\dagger \Delta_k^\dagger \Delta_l \phi +
g \tr(\Delta_1^\dagger\Delta_2) \tr(\Delta_2^\dagger\Delta_1) + 
g' \tr(\Delta_1^\dagger\Delta_1) \tr(\Delta_2^\dagger\Delta_2) 
\nonumber \\
&&
+ \left( t_k\, \phi^\dagger \Delta_k \tilde{\phi} +
\mbox{H.c.} \right),
\label{pot}
\end{eqnarray}
where summation over $k,l=1,2$ is understood.
This potential is not the most general one, as we have omitted some of
the quartic terms. This is justified in view of the scope of this
paper, as laid out in the introduction. Moreover, due to the smallness
of the triplet VEVs, the quartic terms are not important numerically
for the mass matrices of the scalars. 

As in the case with a singlet triplet, we illustrate our main point
here taking all the VEVs $v, w_1, w_2$ as real and positive,
and with real values for $t_1, t_2$  as well.  
Again, the following orders of
magnitude for the parameters in the potential are assumed:
\begin{equation}\label{oomagn2}
a,\: b_{kl} \sim v^2; \quad
c,\: d_{kl},\: e_{kl},\: h_{kl},\: f_{kl},\: g,\: g' \sim 1; \quad
|t_k| \ll v. 
\end{equation}
We also confine ourselves to cases where  $w_1, w_2 \ll v$, keeping in mind
the constraint on the $\rho$-parameter.

In general, the scalar potential~(\ref{pot}) can only be treated numerically.
However, since the triplet VEVs
$w_k$ are small (we will have $w_k \lesssim 1$\,GeV in our
numerical part), it should be a good approximation to 
drop the quartic terms in the scalar triplets.
In the following we will discuss the VEVs and the mass matrices of the doubly
and singly-charged scalars in this approximation, so that our broad conclusions 
are transparent. However, the numerical results 
presented in section~\ref{benchmark} are obtained using the full
potential~(\ref{pot}), including even the effects of the small triplet
VEVs. We find that the results are in very good accordance with the
approximation. 

For the sake of a convenient notation we define the following $2 \times 2$
matrices and vectors: 
\begin{equation} 
B = (b_{kl}), \quad E = (e_{kl}), \quad H = (h_{kl}), \quad
t = \left( \begin{array}{c} t_1 \\ t_2 \end{array} \right), \quad
w = \left( \begin{array}{c} w_1 \\ w_2 \end{array} \right).
\end{equation}
With this notation the conditions for a stationary point of the
potential are given by
\begin{eqnarray}
\left( B + \frac{v^2}{2} \left( E-H \right) \right) w + 
v^2\, t &=& 0, 
\label{vev1} \\
a + cv^2 + \frac{1}{2} w^T (E-H) w + 2\, t \cdot w &=& 0,
\label{vev2}
\end{eqnarray}
where we have used the notation $t \cdot w = \sum_k t_k w_k$.
These two equations are exact if one neglects all terms quartic in the
triplet VEVs in 
$V_0 \equiv V( \langle \phi \rangle_0, \langle \Delta \rangle_0)$.
In equation~(\ref{vev2}) we have already divided by $v$, assuming 
$v \neq 0$. Using equation~(\ref{vev1}), the small VEVs $w_k$
are obtained as
\begin{equation}
w = -v^2 \left( B + \frac{1}{2} v^2 (E-H) \right)^{-1} t.
\end{equation}

Now we discuss the mass matrices of the charged scalars.
A glance at the scalar potential equation~(\ref{pot})---neglecting quartic
terms in the triplet scalars---reveals that the
first two lines of $V$ make no difference between the singly and
doubly-charged scalars. Thus, the difference in the respective mass
matrices originates in the terms of the third line. The mass matrix of
the doubly-charged scalars is obtained as
\begin{equation}
\mm^2_{++} = B + \frac{v^2}{2} \left( E+H \right).
\end{equation}
As for the singly-charged fields $\Delta^+_k$, one has to take into
account that they can mix with $\phi^+$ of the Higgs doublet. Writing
the mass term as 
\begin{equation}\label{m++}
-\mathcal{L}^{\pm}_S = 
\left( \delta^-_1, \delta^-_2, \phi^- \right) \mm^2_+
\left( \begin{array}{c}
\delta^+_1 \\ \delta^+_2 \\ \phi^+ \end{array} \right)
+ \mbox{H.c.},
\end{equation}
equation~(\ref{pot}) leads to 
\begin{equation}\label{m+}
\renewcommand{\arraystretch}{1.2}
\mm^2_+ = \left(
\begin{array}{cc}
B + \frac{v^2}{2} \, E & 
\sqrt{2} v \left( t - H w/2 \right) \\
\sqrt{2} v \left( t - H w/2 \right)^\dagger &
a + cv^2 + \frac{1}{2} w^T (E+H) w
\end{array} \right).
\end{equation}
Obviously, this mass matrix has to have an eigenvector with eigenvalue
zero which corresponds to the would-be-Goldstone boson. Indeed, 
using equations~(\ref{vev1}) and~(\ref{vev2}), we find
\begin{equation}\label{wouldbe}
\mm^2_{+} \left( \begin{array}{c}
v_T \\ v/\sqrt{2} \end{array} \right) = 0,
\end{equation}
which serves as a consistency check.

Note that the matrix $B$ largely controls
the mass of the triplet scalars
and the order of magnitude
of its elements (or of its eigenvalues) is expected to be a little
above the electroweak 
scale, represented 
by $v \simeq 246$\,GeV. On the other
hand, the quantities $t_k$ trigger the small triplet VEVs, so they
should be considerably smaller than the electroweak scale. Therefore, 
in a rough approximation one could neglect the $t_k$ and the triplet VEVs in
the mass matrix $\mm^2_+$. In that limit, also $a + cv^2 = 0$ and 
the charged would-be-Goldstone boson consists entirely of $\phi^+$, without
mixing with the $\delta^+_k$. 

The mass matrices~(\ref{m++}) and~(\ref{m+}) are diagonalized by
\begin{equation}\label{UV}
U^\dagger \mm^2_{++} U = \diag (M^2_1, M^2_2) 
\quad \mbox{and} \quad
V^\dagger \mm^2_{+} V = \diag (\mu_1^2, \mu_2^2, 0),
\end{equation}
respectively, with
\begin{equation}\label{UV1}
\left( \begin{array}{c} \delta^{++}_1 \\ \delta^{++}_2
\end{array} \right) = U
\left( \begin{array}{c} H^{++}_1 \\ H^{++}_2
\end{array} \right), 
\quad
\left( \begin{array}{c} \delta^{+}_1 \\ \delta^{+}_2 \\ \phi^+
\end{array} \right) = V
\left( \begin{array}{c} H^{+}_1 \\ H^{+}_2 \\ G^+
\end{array} \right).
\end{equation}
We have denoted the fields with definite mass by $H^{++}_k$ and
$H^{+}_k$, and $G^+$ is the charged would-be-Goldstone boson.

The gauge Lagrangian relevant for the decays considered in this paper is given
by 
\begin{eqnarray}
\mathcal{L}_\mathrm{gauge} &=& ig \sum_{k=1}^2 \left[
\delta^-_k (\partial^\mu \delta^{++}_k) -
(\partial^\mu \delta^-_k)\, \delta^{++}_k  
\right] W_\mu^- 
\nonumber \\ 
&& 
-\frac{g^2}{\sqrt{2}} \sum_{k=1}^2 w_k W^-_\mu {W^-}^\mu \delta_k^{++}
+ \mbox{H.c.} 
\end{eqnarray}
Here $g$ is the $SU(2)$ gauge coupling constant.
Inserting equation~(\ref{UV1}) into this Lagrangian 
allows us to compute the decay rates of 
$H_1^{++} \to H_2^+ W^+$ and $H_k^{++} \to W^+ W^+$ ($k=1,2$).
The corresponding formulas are found in appendix~\ref{expressions}.

The $\Delta L = 2$ Yukawa interactions between the triplets and the leptons
are   
\begin{equation}\label{Ll}
\mathcal{L}_Y = \frac{1}{2} \,
\sum_{k=1}^2 h^{(k)}_{ij} L^T_i C^{-1} i \tau_2 \Delta_k L_j + 
\mbox{H.c.},
\end{equation}
where $C$ is the charge conjugation matrix, the $h^{(k)}_{ij}$ 
are the symmetric Yukawa coupling matrices of the triplets
$\Delta_k$, and the $i,j$ are the
summation indices over the three neutrino flavours.\footnote{We
  assume the charged-lepton mass matrix to be already
  diagonal.} The $L_i$
denote the left-handed lepton doublets.

The neutrino mass matrix is generated from equation~(\ref{Ll}) 
when the triplets acquire VEVs:
\begin{equation}
(M_\nu)_{ij}
= h^{(1)}_{ij} w_1 + h^{(2)}_{ij} w_2.
\end{equation}
This connects the Yukawa coupling constants $h^{(1)}_{ij}$, 
$h^{(2)}_{ij}$ and the triplet VEVs $w_1$, $w_2$,   
once the neutrino mass matrix is written down for a particular scenario. 
In our subsequent calculations,  we proceed as follows. First of all, the
neutrino mass eigenvalues 
are fixed according to a particular 
type of mass spectrum.
In this work we illustrate our
points, without any loss 
of generality, by resorting to 
normal hierarchy 
of the neutrino mass spectrum 
and setting the lowest
neutrino mass eigenvalue to zero.  
Furthermore, using the observed central values of the various 
lepton mixing angles, 
the elements of the neutrino mass matrix $M_\nu$ can be found by using
the equation 
\begin{equation}\label{Mnu}
M_\nu = U {\hat{M}}_\nu U^\dagger,
\end{equation} 
where $U$ is the PMNS matrix given by~\cite{pdg} 
\begin{equation}
U = 
\left(\begin{array}{ccc}
c_{12} c_{13} & s_{12} c_{13} & s_{13} e^{-i\delta}  \\ - s_{12} c_{23} -
c_{12} s_{23} s_{13} e^{i\delta}  & c_{12} c_{23} - s_{12} s_{23} s_{13}
e^{i\delta} & s_{23} c_{13}  \\ s_{12} s_{23} - c_{12} c_{23} s_{13}
e^{i\delta} & - c_{12} s_{23} - s_{12} c_{23} s_{13} e^{i\delta} & c_{23}
c_{13} 
\end{array}\right)
\end{equation}
and ${\hat{M}}_\nu$ is the diagonal matrix of the neutrino masses.
In equation~(\ref{Mnu}) we have dropped 
possible Majorana phases.
One can use the  recent global analysis of data to  determine the various
entries of $U$~\cite{fogli}.  
We have taken the phase factor $\delta$ to be zero for simplicity.  Then,
using the central values of 
all angles, including that for $\theta_{13}$ as obtained from the recent 
Daya Bay and RENO experiments~\cite{dayabay,RENO},
the left-hand side of 
equation~(\ref{Ll}) 
is completely known, 
at least in orders of magnitude.
The actual mass matrix thus constructed has some elements at least one order
of magnitude smaller than the others, thus suggesting 
texture zeros. 

For each of the benchmark points used in the next section, $w_1$ and
$w_2$, the VEVs of the two triplets, are determined by values of the
parameters in the scalar potential.  Of course, the coupling matrices
$h^{(1)}$ and $h^{(2)}$ are still indeterminate. In order to evolve
a working principle based on economy of free parameters, 
we fix the Yukawa coupling matrix $h^{(2)}$
by choosing one suitable value for all elements of the 
$\mu$--$\tau$ block and another value, a smaller one, 
for the rest of the matrix.
That fixes all the elements of the other
matrix. Although there is a degree of arbitrariness in such a method,
we emphasize that it does not affect the generality of our conclusions,
so long as we adhere to the wide choice of scenarios adopted in the next
section, including both small and large values of the $\Delta L = 2 $
Yukawa couplings.

\section{Benchmark points and doubly-charged scalar decays} 
\label{benchmark}
Our purpose is to investigate the expected changes in the phenomenology
of doubly-charged scalars when two triplets are present. In general,
the two scalars of this kind, namely, 
$H^{++}_1$ and  $H^{++}_2$ 
can both be produced at the LHC via the Drell-Yan process, which
can have about 10\% enhancement from the two-photon channel. They
will, over a large region of the parameter space, have the following
decays: 
\begin{eqnarray}
H^{++}_1 &\rightarrow& \ell^+_i \ell^+_j, \label{hll} \\
H^{++}_1 &\rightarrow& W^+ W^+, \\
H^{++}_1 &\rightarrow& H^+_2  W^+,  \label{hhw} \\
H^{++}_2 &\rightarrow& \ell^+_i \ell^+_j, \\
H^{++}_2 &\rightarrow& W^+ W^+,
\end{eqnarray}
with $\ell_i, \ell_j = e, \mu, \tau$ in equation~(\ref{hll}). 
As we discussed in section~\ref{1-triplet}, 
in the context of the single-triplet model the decay analogous to
equation~(\ref{hhw}) is practically never allowed, 
unless the masses are very low. On the other hand, mixing
between two triplets opens up situations where the mass separation between
$H^{++}_1$ and  $H^{+}_2$  kinematically allows the transition~(\ref{hhw}).
Denoting the mass of $H_k^{++}$ by $M_k$ and that of 
$H_k^+$ by $\mu_k$ ($k=1,2$) 
and using the convention $M_1 > M_2$ and $\mu_1 > \mu_2$, 
this decay is possible if $M_1 > \mu_2 + m_W$. 
We demonstrate numerically that this can naturally happen, by 
considering three distinct regions of the parameter space and selecting
four benchmark points (BPs) for each region.

We have seen that, in a model with a single triplet, the
doubly-charged Higgs decays into either $\ell^+_i \ell^+_j$ or $W^+
W^+$. The former is controlled by the $\Delta L =2$ coupling constants
$h_{ij}$, while the latter is driven by the triplet VEV $w$. Since
neutrino masses are given by $M_{\nu} = h w$, large ($\simeq 1$)
values of $h_{ij}$ imply a small VEV $w$, and vice versa.  
Accordingly, assuming $h_{ij} \neq 0$,
three regions in the parameter space can be identified, 
where one can have
\begin{enumerate}
\renewcommand{\labelenumi}{\roman{enumi})}
\item
$\Gamma(\Delta^{++} \rightarrow \ell^+_i \ell^+_j ) \ll
\Gamma(\Delta^{++} \rightarrow W^+ W^+ )$, 
\item
$\Gamma(\Delta^{++} \rightarrow \ell^+_i \ell^+_j ) \gg
\Gamma(\Delta^{++} \rightarrow W^+ W^+ )$, 
\item
$\Gamma(\Delta^{++} \rightarrow \ell^+_i \ell^+_j ) \sim
\Gamma(\Delta^{++} \rightarrow W^+ W^+ )$. 
\end{enumerate}
In the context of two
triplets, we choose three different `scenarios' in the same
spirit, with similar relative rates of the two channels 
$H^{++}_k \rightarrow \ell^+_i \ell^+_j$ and
$H^{++}_k \rightarrow W^+ W^+$.
Four BPs are selected for each such scenario 
through the appropriate choice of 
parameters in the scalar potential.
The parameters for each BP are listed in appendix~\ref{bps}.  
The resulting masses of the
various physical scalar states are shown in 
tables~\ref{charged} 
and~\ref{neutral}.  Although our
study focuses mainly on the phenomenology of charged scalars, we also
show the masses of the neutral scalars. It
should be noted that the lightest CP-even neutral scalar, which is
dominated by the doublet, has mass around 125 GeV for each BP.

All the twelve BPs (distributed among the three different scenarios)
have $M_1$ sufficiently above $M_2$ to open up
$H^{++}_1 \rightarrow H^+_2 W^+$.  The branching ratios
in different channels are of course dependent on the specific
BP. We list all the branching ratios for $H^{++}_1$ and $H^{++}_2$
in table~\ref{branching}, together with their pair-production 
cross sections at the LHC with $\sqrt{s} = 14$\,TeV. 

The cross sections and branching ratios have been calculated with the help of
the  package FeynRules (version 1.6.0) \cite{neil,duhr}, 
thus creating a new model file in 
CompHEP (version 2.5.4)~\cite{pukhov}.
CTEQ6L parton distribution functions have been used,
with the renormalisation and factorisation scales set at the doubly-charged
scalar mass. Using the full machinery of scalar mixing in this model,
the decay widths into various channels have been obtained,
for which the relevant expressions are presented in appendix~\ref{expressions}. 

The results summarised in 
table~\ref{branching} 
show that,
for the decay of $H^{++}_1$, the channel 
$H^+_2 W^+$ is dominant for two of the four BPs in scenario~1 and 
all four BPs in scenarios~2 and~3. 
This, in the first place, substantiates
our claim that one may have to look for a singly-charged scalar 
in the final state 
that opens up when more than one doublet is present.
This is because, for the BPs where $H^{++}_1 \rightarrow H^+_2 W^+$
dominates, the branching ratios for the other final states are far
too small to yield any detectable rates.

\section{Usefulness of $H_1^{++} \longrightarrow H_2^{+} W^+$ at the LHC}
\label{usefulness}
Table 3  contains the rates for pair-production of the heavier as well
as the lighter doubly-charged scalar at the 14 TeV run of the LHC. A quick
look at these rates revals that, for the heavier of the doubly-charged 
scalars, it
varies from about 1.4\,fb to 3.6\,fb, for  masses ranging
approximately between 400 and 550\,GeV. 
Therefore, as can be read off from table~\ref{branching},
for ten of our twelve BPs, 
an integrated luminosity of about 
500\,fb$^{-1}$ 
is likely 
to yield about $700$ to $1800$ events of the 
$H^{+}_2 W^{+} H^{-}_2 W^{-}$
type. Keeping in mind the fact that 
$H^{+}_2$ 
mostly decays in the
channel $H^{+}_2 \rightarrow \ell^+ \bar{\nu}_{\ell}$, such final
states should {\it prima facie} be observed at the LHC, although 
event selection strategies of a very special nature may be required
to distinguish the $H^{+}_2$ from a $W^+$ decaying into $\ell^+ \nu_{\ell}$.   

The primary advantage of focusing on the channel  
$H_1^{++} \longrightarrow H_2^{+} W^+$ is that it helps one
in differentiating between the two kinds of type~II 
cases,
namely those containing 
one and 
two scalar triplets, respectively. 
In order to emphasize this point,  
we summarize 
below the result of 
a simulation in the context of the 14 TeV run of the LHC. For our simulation,
the amplitudes have been computed using the package Feynrules (version 1.6.0), 
with the subsequent event generation through MadGraph 
(version 5.12)~\cite{mad},
and showering 
with the help of PYTHIA 8.0. CTEQ6L parton distribution functions
have been used.

We compare the two-triplet case with the single-triplet case. 
In the first case, there are two doubly charged
scalars, and one has contributions from both $H_1^{\pm\pm}$ and
$H_2^{\pm\pm}$ to the leptonic final states following their Drell-Yan
production. While the former, in the chosen benchmark points, decays
into $H_1^{\pm}W^{\pm}$, the latter goes either to a same-sign 
$W$-pair
or 
to 
same-sign dileptons. If one considers two, 
three and
four-lepton final states with missing transverse energy (MET), there
will be contributions from both of the doubly-charged scalars, with
appropriate branching ratios, combinatoric factors and response to the
cuts imposed. We have carried out our analysis with a set of cuts
listed in table~4, which are helpful in suppressing the standard model
backgrounds. Thus one can define the following ratios of events emerging
after the application of cuts:
\begin{equation}
r_1 = \frac{\sigma(4\ell + \mathrm{MET})}{\sigma(3\ell + \mathrm{MET})} ,
\qquad
r_2 = \frac{\sigma(4\ell + \mathrm{MET})}{\sigma(2\ell + \mathrm{MET})}.
\end{equation}

The values of these ratios for the 
%%%%
three scenarios of BP~3 
are presented in table~5.
In each case, the ratios for the 
two-triplet case is
presented alongside the corresponding  
single-triplet case, 
with the
mass of the doubly charged scalar in the latter case being close to
that of the lighter state $H_2^{\pm\pm}$ in the former. Both of the
situations where, in the later case, the doubly charged scalar decays
dominantly into either $W^{\pm}W^{\pm}$ or $\ell^{\pm}\ell^{\pm}$ are
represented in our illustrative results. One can clearly notice from
the results (which apply largely to our other benchmark points as well)
that both $r_1$ and $r_2$ remain substantially larger in the 
two-triplet case 
as compared to the  
single-triplet case. 
One reason for this is  
an enhancement via the combinatoric factors in the two-triplet case.
However, the more important reason is that the 4$\ell$ events survive
the MET cut with greater efficiency. In the single-triplet case, the
survival rate efficiency is extremely small when $H^{\pm\pm}$ decays
mainly into same-sign dileptons, the MET coming mostly from energy-momentum
mismeasurement (as a result of lepton energy smearing) or 
initial and final-state radiation. 
In the two-triplet case, on the other hand, 
the decay  $H_1^{++} \longrightarrow H_2^{+} W^+$ leaves ample scope
for having MET in  
$W$-decays 
as well as in the decay 
$H_2^+ \longrightarrow \ell^+ \bar\nu_\ell$, 
thus leading to substantially
higher cut survival efficiency. Thus, from an examination of
such numbers as those presented in table 5, one can quite effectively
use the channel $H_1^{++} \longrightarrow H_2^{+} W^+$ to 
distinguish a two-triplet case from a single-triplet case, provided 
the heavier doubly-charged state is within the kinematic reach of the LHC.

\begin{table}%[h]
\centering  % centering table
\begin{tabular}{|c||c|c|c|c|c|}

  \hline
   & Mass (GeV) & BP 1 & BP 2 & BP 3 & BP 4 \\ %1st row
   \hline \hline
   & $H_1^{++}$ & $515.99$ & $515.99$ & $521.54$ & $524.15$
   \\ \cline{2-6} %2nd row 
  Scenario 1 &  $H_2^{++}$ & $443.04$ & $429.16$ & $455.59$ &
   $470.15$ \\ \cline{2-6} %2nd row\\ 
   & $H_1^+$ & $515.98$ & $515.98$ & $498.97$ & $515.78$ \\ \cline{2-6} %2nd row
   & $H_2^+$ & $368.45$ & $360.15$ & $423.26$ & $418.65$ \\ 
  \hline
     & $H_1^{++}$ & $526.78$ & $525.00$ & $429.13$ & $464.31$
  \\ \cline{2-6} %2nd row 
  Scenario 2 &  $H_2^{++}$ & $414.18$ & $401.63$ & $392.45$ &
  $407.20$ \\ \cline{2-6} %2nd row\\ 
   & $H_1^+$ & $520.26$ & $519.86$ & $414.48$ & $459.23$ \\ \cline{2-6} %2nd row
   & $H_2^+$ & $343.28$ & $334.97$ & $339.02$ & $340.63$ \\ 
  \hline
     & $H_1^{++}$ & $521.54$ & $464.31$ & $525.00$ & $429.13$
  \\ \cline{2-6} %2nd row
  Scenario 3 &  $H_2^{++}$ & $455.59$ & $407.20$ & $401.63$ &
  $392.45$ \\ \cline{2-6} %2nd row\\ 
   & $H_1^+$ & $498.97$ & $459.23$ & $519.86$ & $414.48$ \\ \cline{2-6} %2nd row
   & $H_2^+$ & $423.26$ & $340.63$ & $334.97$ & $339.02$ \\ 
  \hline
\end{tabular}
\caption{Charged scalar masses. \label{charged}}  
\end{table}
%
%\newpage
%
\begin{table}%[h]
\centering  % centering table
\begin{tabular}{|c||c|c|c|c|c|}

  \hline
   & Mass (GeV) & BP 1 & BP 2 & BP 3 & BP 4 \\ %1st row
   \hline \hline
   & $H_{1R}^0$ & $365.70$ & $364.86$ & $350.39$ & $364.59$
   \\ \cline{2-6} %2nd row 
   & $H_{2R}^0$ & $193.89$ & $194.00$ & $256.09$ & $245.96$
   \\ \cline{2-6} %2nd row 
  Scenario 1 &   $H_{3R}^0$ & $125.00$ & $125.03$ & $125.01$ &
   $125.01$ \\ \cline{2-6} %2nd row 
   &  $H_{1I}^0$ & $364.98$ & $364.85$ & $350.39$ & $364.59$
   \\ \cline{2-6} %2nd row 
   &  $H_{2I}^0$ & $194.43$ & $193.98$ & $256.08$ & $245.96$
   \\ \cline{2-6} %2nd row \\  
  \hline
   & $H_{1R}^0$ & $365.69$ & $365.70$ & $295.58$ & $325.51$
  \\ \cline{2-6} %2nd row 
   & $H_{2R}^0$ & $173.97$ & $173.96$ & $173.98$ & $173.96$
  \\ \cline{2-6} %2nd row 
  Scenario 2 &   $H_{3R}^0$ & $125.02$ & $125.02$ & $125.04$ &
  $125.02$ \\ \cline{2-6} %2nd row 
   &  $H_{1I}^0$ & $365.69$ & $365.70$ & $295.59$ & $325.52$
  \\ \cline{2-6} %2nd row 
   &  $H_{2I}^0$ & $173.97$ & $173.96$ & $173.98$ & $173.96$
  \\ \cline{2-6} %2nd row  
  \hline
   & $H_{1R}^0$ & $350.39$ & $325.51$ & $365.69$ & $295.58$
  \\ \cline{2-6} %2nd row 
   & $H_{2R}^0$ & $256.08$ & $173.96$ & $173.98$ & $173.96$
  \\ \cline{2-6} %2nd row 
  Scenario 3 &   $H_{3R}^0$ & $125.02$ & $125.02$ & $125.04$ &
  $125.02$ \\ \cline{2-6} %2nd row 
   &  $H_{1I}^0$ & $350.39$ & $325.51$ & $365.69$ & $295.58$
  \\ \cline{2-6} %2nd row 
   &  $H_{2I}^0$ & $256.08$ & $173.96$ & $173.98$ & $173.96$
  \\ \cline{2-6} %2nd row 
  \hline
\end{tabular}
\caption{Neutral scalar masses. \label{neutral}} 
\end{table}
%
%\newpage
%
%
\begin{table}%[h]
\centering  % centering table
\begin{tabular}{|c||c|c|c|c|c|}
  \hline
   & Data & BP 1 & BP 2 & BP 3 & BP 4 \\ %1st row
   \hline \hline
 & $\mbox{BR}(H_1^{++} \rightarrow H_2^+ W^+)$&$0.08 $
   &$0.10$&$0.99$&$0.99$\\ \cline{2-6} %2nd row  
& $\mbox{BR}(H_1^{++} \rightarrow W^+
   W^+)$&$0.92$&$0.90$&$0.01$&$0.004$\\ \cline{2-6} %2nd row 
& $\mbox{BR}(H_1^{++} \rightarrow \ell^+_i \ell^+_j)$&$3.89 \times 10^{-17}$&$3.82
   \times 10^{-17}$&$3.34 \times 10^{-20}$&$8.044 \times
   10^{-21}$\\ \cline{2-6} %2nd row 
Scenario 1 & $\mbox{BR}(H_2^{++} \rightarrow W^+
   W^+)$&$0.99$&$0.99$&$0.99$&$0.99$\\ \cline{2-6} %2nd row 
& $\mbox{BR}(H_2^{++} \rightarrow \ell^+_i \ell^+_j)$&$1.76 \times 10^{-20}$&$1.72
   \times 10^{-20}$&$1.78 \times 10^{-18}$&$1.76 \times
   10^{-19}$\\ \cline{2-6} %2nd row 
 & $\sigma(pp \rightarrow H_1^{++} H_1^{--})$&$\ 1.664$\,fb
&$\ 1.534$\,fb&$\ 1.446$\,fb&$\ 1.408$\,fb\\ \cline{2-6}
   %2nd row 
& $\sigma(pp \rightarrow H_2^{++} H_2^{--})$&$\ 3.044$\,fb&$\ 3.5
  $\,fb&$\ 2.714$\,fb&$\ 2.308$\,fb\\ 
   \hline
 & $\mbox{BR}(H_1^{++} \rightarrow H_2^+
   W^+)$&$0.99$&$0.99$&$0.99$&$0.98$\\ \cline{2-6} %2nd row  
 & $\mbox{BR}(H_1^{++} \rightarrow W^+ W^+)$&$7.44 \times 10^{-22}$&$6.67
   \times 10^{-22}$&$1.08 \times 10^{-18}$&$1.77 \times
   10^{-21}$\\ \cline{2-6} %2nd row  
 & $\mbox{BR}(H_1^{++} \rightarrow \ell^+_i \ell^+_j)$ 
&$0.01$&$0.01$&$0.001$&$0.02$\\ \cline{2-6} %2nd row  
Scenario 2 & $\mbox{BR}(H_2^{++} \rightarrow W^+ W^+)$&$3.75 \times
10^{-19}$&$3.39 \times 10^{-19}$&$8.28 \times 10^{-15}$&$4.16 \times
10^{-19}$\\ \cline{2-6} %2nd row 
& $\mbox{BR}(H_2^{++} \rightarrow \ell^+_i \ell^+_j)$
&$0.99$&$0.99$&$0.99$&$0.99$\\ \cline{2-6} %2nd row  
& $\sigma(pp \rightarrow H_1^{++} H_1^{--})$&$\ 1.36$\,fb&
$\ 1.41$\,fb&$\ 3.59$\,fb&$\ 2.46$\,fb\\ \cline{2-6} %2nd row  
& $\sigma(pp \rightarrow H_2^{++} H_2^{--})$&$\ 3.98$\,fb&
$\ 4.65$\,fb&$\ 5.28$\,fb&$\ 4.38$\,fb\\ 
  \hline
& $\mbox{BR}(H_1^{++} \rightarrow H_2^+
  W^+)$&$0.99$&$0.99$&$0.99$&$0.99$\\ \cline{2-6} %2nd row 
& $\mbox{BR}(H_1^{++} \rightarrow W^+ W^+)$&$5.56 \times 10^{-13}$&$1.79
  \times 10^{-11}$&$6.75 \times 10^{-12}$&$1.1 \times
  10^{-10}$\\ \cline{2-6} %2nd row 
& $\mbox{BR}(H_1^{++} \rightarrow \ell^+_i \ell^+_j)$&$3.69 \times 10^{-10}$&$1.26
  \times 10^{-12}$&$1.16 \times 10^{-12}$&$5.48 \times
  10^{-12}$\\ \cline{2-6} %2nd row 
Scenario 3 & $\mbox{BR}(H_2^{++} \rightarrow W^+ W^+)$&$0.0001
$&$0.98$&$0.97$&$0.99$\\ \cline{2-6} %2nd row 
& $\mbox{BR}(H_2^{++} \rightarrow \ell^+_i \ell^+_j)$
&$0.99$&$0.02$&$0.03$&$0.01$\\ \cline{2-6} %2nd row 
& $\sigma(pp \rightarrow H_1^{++} H_1^{--})$&$\ 1.45$\,fb&
$\ 2.46$\,fb&$\ 1.41$\,fb&$\ 3.59$\,fb\\ \cline{2-6} %2nd row 
& $\sigma(pp \rightarrow H_2^{++} H_2^{--})$&$\ 2.71$\,fb&
$\ 4.38$\,fb&$\ 4.65$\,fb&$\ 5.28$\,fb\\ 
  \hline
\end{tabular}
\caption{Decay branching ratios and production cross sections for
  doubly-charged scalars. \label{branching}} 
\end{table}

\begin{table}[h]
\centering  % centering table
%\begin{table}[htb!]
%\centering
\begin{tabular}{|l|}
\hline
$\mathrm{MET}  > 70$ GeV \\
\hline
$ \Sigma |p_T^\mathrm{vis}| + \mathrm{MET}  > 500$ GeV \\
\hline
$\left| p_T^\mathrm{lepton} \right| > 30$ GeV \\
\hline
$|\eta_{\,\mathrm{lep}}| < 2.5$ \\
\hline
$|\eta_{\,\mathrm{jet}}| < 4.5$ \\
\hline
\end{tabular}
\caption{Cuts used for determination of ratios of events $r_1$ and $r_2$.
The subscript $T$ stands for `transverse' and $\eta$ denotes the
pseudorapidity.} 
\end{table}
\begin{table}%[h]
\centering  % centering table
\begin{tabular}{|c||c|c|c|}
  \hline
BP 3 & Ratio & Two triplets & One triplet \\ %1st row
   \hline \hline
\raisebox{-10pt}[0pt][0pt]{Scenario 1} & $r_1$ & $0.20$ & $0.04$ \\ \cline{2-4}   
           & $r_2$ & $0.05$ & $0.01$ \\ \cline{2-4} 
    \hline 
\raisebox{-10pt}[0pt][0pt]{Scenario 2} & $r_1$ & $0.44$ & $ < 10^{-6}$ \\ \cline{2-4}   
           & $r_2$ & $0.21$ & $ < 10^{-9}$ \\ \cline{2-4} 
    \hline  
\raisebox{-10pt}[0pt][0pt]{Scenario 3} & $r_1$ & $0.12$ & $ < 10^{-5} $ \\ \cline{2-4}   
           & $r_2$ & $0.04$ & $ < 10^{-6}$ \\ \cline{2-4} 
    \hline

\end{tabular}
\caption{Ratio of events $r_1$, $r_2$ for two-triplet and single-triplet scenario respectively for benchmark point 3.} 
\end{table}

\section{Summary and conclusions}
\label{concl}
In this paper, we have argued, 
taking models with the type~II seesaw
mechanism for neutrino mass generation as a motivation, 
that it makes sense to consider 
scenarios with more than one scalar triplet. As the simplest
extension, we have formulated in detail a model with two $Y = 2$
complex triplets of this kind. On taking into account the mixing of
the triplets with each other (and also with the doublet, albeit with
considerable restriction), and thus identifying all the mass eigenstates
along with their various interaction strengths, we find that
the heavier doubly-charged scalar decays dominantly into 
the lighter singly-charged scalar and a $W$ boson over a large
region of the parameter space. It should be re-iterated that this feature is a
generic one and is avoided only  
in very limited situations 
or in the case of unusually high values 
of the triplet Yukawa coupling. The deciding factor here is the
decay being driven  
by the $SU(2)$ gauge coupling. 

Thus the above mode is often 
the only way of looking for the heavier doubly-charged scalar state
and thus for the existence of two scalar triplets. 
Our choice of benchmark points
for reaching this conclusion spans cases where the $\Delta L = 2$
lepton couplings of the triplets have values at the high (close to one)
and low as well as the intermediate level, 
consistent with 
the observed neutrino
mass and mixing patterns. In general, with the heavier triplet mass 
ranging up to more than 500\,GeV, one expects about 700 to 1800 events
of the type 
$pp \rightarrow H^{+}_2 W^{+} H^{-}_2 W^{-}$ 
at the
14\,TeV run of the LHC, for an integrated luminosity of 500\,fb$^{-1}$. We
have also demonstrated that ratios of the numbers of two, three and
four-lepton events with MET offer a rather spectacular distinction of the
two-triplet case 
from one with a single triplet only. 
It is thus both interesting and challenging to look for this mode,
with well-defined criteria for distinguishing the 
$H^{+}_2$ through its decay products.

\paragraph{Acknowledgements:}
The work of A.C.\ and B.M.\ has been
partially supported by the Department of Atomic Energy, 
Government of India, through funding available for
the Regional Centre for Accelerator-Based Particle Physics, 
Harish-Chandra Research Institute. 
B.M.\ acknowledges the
hospitality of the Faculty of Physics, University of Vienna,
at the formative stage of this project.
A.C.\ thanks AseshKrishna Dutta, Tanumoy Mandal, Kenji Nishiwaki and 
Saurabh Niyogi for many helpful discussions.

\appendix

\section{Input parameters for the various benchmark points}
\label{bps}
For the definition of the parameters of the scalar potential see
equation~(\ref{pot}). The parameter $a$ and the elements of the matrix
$B$ are in units of GeV$^2$, the $t_k$ are in units of GeV, while all
other parameters of the potential are dimensionless.
The Yukawa coupling matrices are defined in equation~(\ref{Ll}).
\subsection{Input parameters for Scenario 1: \\[-1mm]
$\mathrm{BR}(H_1^{++} \rightarrow W^+ W^+) \gg 
\mathrm{BR}(H_1^{++} \rightarrow \ell_i \ell_j)$ }
\paragraph{BP 1:} 
The input parameters for the scalar potential are
\[
a = -15625, \quad
\frac{1}{2}\,B = \left(
\begin{array}{cc}
60508 & -74990 \\ -74990 & 60591.2
\end{array} \right),
\]
\[
\frac{1}{4}\,D =  \left(
\begin{array}{cc}
1 & 0.89 \\ 0.89 & 1
\end{array} \right),
\;
\frac{1}{2}\,(E-H) = \left(
\begin{array}{cc} 0.82 & 0.9 \\ 0.9 & 0.82
\end{array} \right), 
\;
H = \left( \begin{array}{cc} 1 & 1 \\ 1 & 1 
\end{array} \right), 
\;
\frac{1}{4}\,F = \left( \begin{array}{cc} 1 & 0.5 \\ 0.5 & 1 
\end{array} \right)
\]
and 
\[
c = 0.26, \; g = g' = 0.89, \; t_1 = -1, \; t_2 = -2.
\]

For these parameter values, the VEVs obtained from
minimization conditions are 
$v = 246.02$\,GeV, $w_1 = 1.09$\,GeV, $w_2 = 1.32$\,GeV. 

The Yukawa coupling matrices are fixed to be
\begin{eqnarray*}
h^{(1)}_{ij} &=&  
\left(\begin{array}{ccc}
2.25 \times 10^{-12} & 5.70 \times 10^{-12} & 1.62 \times 10^{-12}
\\ 
5.70 \times 10^{- 12}  & 0.80 \times 10^{-11}  & 0.66 \times 10^{-11} 
\\ 
1.62 \times 10^{- 12}  & 0.66 \times 10^{- 11} & 1.74 \times 10^{- 11}
\end{array}\right),
\\
h^{(2)}_{ij} &=& 
\left(\begin{array}{ccc}
1.0 \times 10^{-12} & 1.0 \times 10^{- 12} & 1.0 \times 10^{- 12}
\\ 1.0 \times 10^{- 12}  & 1.0 \times 10^{-11} & 1.0 \times 10^{-11}  
\\ 
1.0 \times 10^{- 12}  & 1.0 \times 10^{-11} & 1.0 \times 10^{-11} 
\end{array}\right).
\end{eqnarray*}
\paragraph{BP 2:} 
The input parameters for the scalar potential are
\[
a = -15625, \quad
\frac{1}{2}\,B = \left(
\begin{array}{cc}
60509.6 & -74990 \\ -74990 & 60590
\end{array} \right),
\]
\[
\frac{1}{4}\,D =  \left(
\begin{array}{cc}
1 & 0.9 \\ 0.9 & 1
\end{array} \right),
\;
\frac{1}{2}\,(E-H) = \left(
\begin{array}{cc} 0.82 & 0.9 \\ 0.9 & 0.82
\end{array} \right), 
\;
H = \left( \begin{array}{cc} 0.9 & 0.9 \\ 0.9 & 0.9 
\end{array} \right), 
\;
\frac{1}{4}\,F = \left( \begin{array}{cc} 0.9 & 0.45 \\ 0.45 & 0.9 
\end{array} \right)
\]
and 
\[
c = 0.26, \; g = g' = 0.9, \; t_1 = -1, \; t_2 = -2.
\]

For these parameter values, the VEVs obtained from
minimization conditions are 
$v = 246.02$\,GeV, $w_1 = 1.09$\,GeV, $w_2 = 1.32$\,GeV. 

The Yukawa coupling matrices are fixed to be 
\begin{eqnarray*}
h^{(1)}_{ij} &=&
\left(\begin{array}{ccc}
2.25 \times 10^{-12} & 5.69 \times 10^{- 12} & 1.62 \times 10^{- 12}
\\ 5.69 \times 10^{- 12}  & 0.79 \times 10^{- 11}  & 0.66 \times 10^{-
  11}  \\ 1.62 \times 10^{- 12}  & 0.66 \times 10^{- 11} & 1.74 \times
10^{- 11} 
\end{array}\right),
\\
h^{(2)}_{ij} &=& 
\left(\begin{array}{ccc}
1.0 \times 10^{-12} & 1.0 \times 10^{- 12} & 1.0 \times 10^{- 12}
\\ 
1.0 \times 10^{- 12}  & 1.0 \times 10^{-11}  & 1.0 \times 10^{-11}  
\\ 
1.0 \times 10^{- 12}  & 1.0 \times 10^{- 11} & 1.0 \times
10^{- 11} 
\end{array}\right).
\end{eqnarray*}
\paragraph{BP 3:}
The input parameters for the scalar potential are
\[
a = -15625, \quad
\frac{1}{2}\,B = \left(
\begin{array}{cc}
58870 & -55110 \\ -55110 & 75000
\end{array} \right),
\]
\[
\frac{1}{4}\,D =  \left(
\begin{array}{cc}
1 & 1 \\ 1 & 1
\end{array} \right),
\;
\frac{1}{2}\,(E-H) = \left(
\begin{array}{cc} 0.8 & 0.95 \\ 0.95 & 1
\end{array} \right), 
\;
H = \left( \begin{array}{cc} 0.7 & 1 \\ 1 & 1 
\end{array} \right), 
\;
\frac{1}{4}\,F = \left( \begin{array}{cc} 0.9 & 0.5 \\ 0.5 & 0.9 
\end{array} \right)
\]
and 
\[
c = 0.2582, \; g = g' = 1, \; t_1 = -1, \; t_2 = -2.
\]

For these parameter values, the VEVs obtained from
minimization conditions are 
$v = 246.02$\,GeV, $w_1 = 0.59$\,GeV, $w_2 = 0.72$\,GeV. 

The Yukawa coupling matrices are fixed to be
\begin{eqnarray*}
h^{(1)}_{ij} &=&
\left(\begin{array}{ccc}
5.15 \times 10^{-12} & 1.15 \times 10^{-11} & 3.98 \times 10^{-12}  
\\ 
1.15 \times 10^{-11}  & 2.48 \times 10^{-11} &  2.22 \times 10^{-11}  
\\ 3.98 \times 10^{- 12}  & 2.22 \times 10^{- 11} & 4.21 \times 10^{- 11}
\end{array}\right),
\\
h^{(2)}_{ij} &=&
\left(\begin{array}{ccc}
1.0 \times 10^{-12} & 1.0 \times 10^{- 12} & 1.0 \times 10^{- 12}
\\ 
1.0 \times 10^{- 12}  & 1.0 \times 10^{- 11}  & 1.0 \times 10^{-11}  
\\ 
1.0 \times 10^{- 12}  & 1.0 \times 10^{- 11} & 1.0 \times 10^{-11} 
\end{array}\right).
\end{eqnarray*}
\paragraph{BP 4:}
The input parameters for the scalar potential are
\[
a = -15625, \quad
\frac{1}{2}\,B = \left(
\begin{array}{cc}
62945 & -65200 \\ -65200 & 76000
\end{array} \right),
\]
\[
\frac{1}{4}\,D =  \left(
\begin{array}{cc}
1 & 0.9 \\ 0.9 & 1
\end{array} \right),
\;
\frac{1}{2}\,(E-H) = \left(
\begin{array}{cc} 0.8 & 1 \\ 1 & 1
\end{array} \right), 
\;
H = \left( \begin{array}{cc} 0.8 & 1 \\ 1 & 1 
\end{array} \right), 
\;
\frac{1}{4}\,F = \left( \begin{array}{cc} 0.8 & 0.5 \\ 0.5 & 1 
\end{array} \right)
\]
and 
\[
c = 0.2582, \; g = g' = 0.9, \; t_1 = -1, \; t_2 = -2.
\]
For these parameter values, the VEVs obtained from
minimization conditions are 
$v = 246.02$\,GeV, $w_1 = 0.66$\,GeV, $w_2 = 0.79$\,GeV. 

The Yukawa coupling matrices are fixed to be 
\begin{eqnarray*}
h^{(1)}_{ij} &=&
\left(\begin{array}{ccc}
4.52 \times 10^{-12} & 1.02 \times 10^{-11} & 3.47 \times 10^{-12} 
\\ 
1.02 \times 10^{- 11}  & 2.12 \times 10^{-11} & 1.90 \times 10^{-11}  
\\ 
3.47 \times 10^{- 12}  & 1.90 \times 10^{- 11} & 3.68 \times 10^{- 11}
\end{array}\right),
\\
h^{(2)}_{ij} &=& 
\left(\begin{array}{ccc}
1.0 \times 10^{-12} & 1.0 \times 10^{-12} & 1.0 \times 10^{-12} 
\\ 
1.0 \times 10^{- 12}  & 1.0 \times 10^{-11}  & 1.0 \times 10^{-11} 
\\ 
1.0 \times 10^{- 12}  & 1.0 \times 10^{- 11} & 1.0 \times 10^{- 11}
\end{array}\right).
\end{eqnarray*}
\subsection{Input parameters for Scenario 2: \\[-1mm]
$\mathrm{BR}(H_1^{++} \rightarrow W^+ W^+) \ll 
\mathrm{BR}(H_1^{++} \rightarrow \ell_i \ell_j)$}
\paragraph{BP 1:}
The input parameters for the scalar potential are
\[
a = -15627, \quad
\frac{1}{2}\,B = \left(
\begin{array}{cc}
77079.1 & -74990 \\ -74990 & 37283.5
\end{array} \right),
\]
\[
\frac{1}{4}\,D =  \left(
\begin{array}{cc}
1 & 0.89 \\ 0.89 & 1
\end{array} \right),
\;
\frac{1}{2}\,(E-H) = \left(
\begin{array}{cc} 0.82 & 0.9 \\ 0.9 & 0.82
\end{array} \right), 
\;
H = \left( \begin{array}{cc} 1 & 1 \\ 1 & 1 
\end{array} \right), 
\;
\frac{1}{4}\,F = \left( \begin{array}{cc} 1 & 0.5 \\ 0.5 & 1 
\end{array} \right)
\]
and 
\[
c = 0.2582, \; g = g' = 0.89, \; t_1 = -1 \times 10^{-9}, \; 
t_2 = -1.5 \times 10^{-9}.
\]
For these parameter values, the VEVs obtained from
minimization conditions are 
$v = 246.01$\,GeV, $w_1 = 1.0 \times 10^{-9}$\,GeV, 
$w_2 = 1.5 \times 10^{-9}$\,GeV. 

The Yukawa coupling matrices are fixed to be 
\begin{eqnarray*}
h^{(1)}_{ij} &=& 
\left(\begin{array}{ccc}
7.84 \times 10^{-4} & 4.55 \times 10^{- 3} & 8.71 \times 10^{-5}  
\\ 
4.55 \times 10^{- 3}  & 1.19 \times 10^{- 2}  & 1.05 \times 10^{- 2} 
\\ 
8.71 \times 10^{- 5}  & 1.05 \times 10^{- 2} & 2.22 \times 10^{- 2}
\end{array}\right),
\\
h^{(2)}_{ij} &=&
\left(\begin{array}{ccc}
2.0 \times 10^{-3} & 2.0 \times 10^{-3} & 2.0 \times 10^{-3} 
\\ 
2.0 \times 10^{-3}  & 1.0 \times 10^{-2} & 1.0 \times 10^{-2} 
\\ 
2.0 \times 10^{-3}  & 1.0 \times 10^{-2} & 1.0 \times 10^{-2}
\end{array}\right).
\end{eqnarray*}
\paragraph{BP 2:}

The input parameters for the scalar potential are
\[
a = -15627, \quad
\frac{1}{2}\,B = \left(
\begin{array}{cc}
77079.1 & -74990 \\ -74990 & 37283.5
\end{array} \right),
\]
\[
\frac{1}{4}\,D =  \left(
\begin{array}{cc}
1 & 0.9 \\ 0.9 & 1
\end{array} \right),
\;
\frac{1}{2}\,(E-H) = \left(
\begin{array}{cc} 0.82 & 0.9 \\ 0.9 & 0.82
\end{array} \right), 
\;
H = \left( \begin{array}{cc} 0.9 & 0.9 \\ 0.9 & 0.9 
\end{array} \right), 
\;
\frac{1}{4}\,F = \left( \begin{array}{cc} 0.9 & 0.45 \\ 0.45 & 0.9 
\end{array} \right)
\]
and 
\[
c = 0.2582, \; g = g' = 0.9, \; t_1 = -1 \times 10^{-9}, \; 
t_2 = -1.5 \times 10^{-9}.
\]
For these parameter values, the VEVs obtained from
minimization conditions are 
$v = 246.01$\,GeV, $w_1 = 1.0 \times 10^{-9}$\,GeV, 
$w_2 = 1.5 \times 10^{-9}$\,GeV. 

The Yukawa coupling matrices are fixed to be
\begin{eqnarray*}
h^{(1)}_{ij} &=& 
\left(\begin{array}{ccc}
7.84 \times 10^{-4} & 4.55 \times 10^{- 3} & 8.71 \times 10^{- 5} 
\\ 
4.55 \times 10^{- 3}  & 1.19 \times 10^{- 2}  & 1.05 \times 10^{- 2} 
\\ 
8.71 \times 10^{- 5}  & 1.05 \times 10^{- 2} & 2.22 \times 10^{- 2}
\end{array}\right),
\\
h^{(2)}_{ij} &=&
\left(\begin{array}{ccc}
2.0 \times 10^{-3} & 2.0 \times 10^{-3} & 2.0 \times 10^{-3} 
\\ 
2.0 \times 10^{-3}  & 1.0 \times 10^{-2}  & 1.0 \times 10^{-2} 
\\ 
2.0 \times 10^{-3}  & 1.0 \times 10^{-2} & 1.0 \times 10^{-2}
\end{array}\right).
\end{eqnarray*}
\paragraph{BP 3:}

The input parameters for the scalar potential are
\[
a = -15627, \quad
\frac{1}{2}\,B = \left(
\begin{array}{cc}
45594.7 & -55110 \\ -55110 & 17574.4
\end{array} \right),
\]
\[
\frac{1}{4}\,D =  \left(
\begin{array}{cc}
1 & 1 \\ 1 & 1
\end{array} \right),
\;
\frac{1}{2}\,(E-H) = \left(
\begin{array}{cc} 0.8 & 0.95 \\ 0.95 & 1
\end{array} \right), 
\;
H = \left( \begin{array}{cc} 0.7 & 1 \\ 1 & 1 
\end{array} \right), 
\;
\frac{1}{4}\,F = \left( \begin{array}{cc} 0.7 & 0.5 \\ 0.5 & 1 
\end{array} \right)
\]
and 
\[
c = 0.2582, \; g = g' = 1, \; t_1 = -1 \times 10^{-8}, \; 
t_2 = -1.5 \times 10^{-8}.
\]
For these parameter values, the VEVs obtained from
minimization conditions are 
$v = 246.01$\,GeV, $w_1 = 1.0 \times 10^{-8}$\,GeV, 
$w_2 = 1.5 \times 10^{-8}$\,GeV. 

The Yukawa coupling matrices are fixed to be
\begin{eqnarray*}
h^{(1)}_{ij} &=& 
\left(\begin{array}{ccc}
7.84 \times 10^{-5} & 4.55 \times 10^{- 4} & 8.71 \times 10^{- 6} 
\\ 
4.55 \times 10^{- 4}  & 1.19 \times 10^{-3}  & 1.05 \times 10^{-3} 
\\ 
8.71 \times 10^{- 6}  & 1.05 \times 10^{- 3} & 2.22 \times 10^{- 3}
\end{array}\right),
\\
h^{(2)}_{ij} &=&
\left(\begin{array}{ccc}
2.0 \times 10^{-4} & 2.0 \times 10^{-4} & 2.0 \times 10^{-4} 
\\ 
2.0 \times 10^{-4}  & 1.0 \times 10^{-3}  & 1.0 \times 10^{-3} 
\\ 
2.0 \times 10^{-4}  & 1.0 \times 10^{-3} & 1.0 \times 10^{-3}
\end{array}\right).
\end{eqnarray*}
\paragraph{BP 4:}

The input parameters for the scalar potential are
\[
a = -15627, \quad
\frac{1}{2}\,B = \left(
\begin{array}{cc}
58460.1 & -65200 \\ -65200 & 23292.4
\end{array} \right),
\]
\[
\frac{1}{4}\,D =  \left(
\begin{array}{cc}
1 & 0.9 \\ 0.9 & 1
\end{array} \right),
\;
\frac{1}{2}\,(E-H) = \left(
\begin{array}{cc} 0.8 & 1 \\ 1 & 1
\end{array} \right), 
\;
H = \left( \begin{array}{cc} 0.8 & 1 \\ 1 & 1 
\end{array} \right), 
\;
\frac{1}{4}\,F = \left( \begin{array}{cc} 0.8 & 0.5 \\ 0.5 & 1 
\end{array} \right)
\]
and 
\[
c = 0.2582, \; g = g' = 0.9, \; t_1 = -1 \times 10^{-9}, \; 
t_2 = -1.5 \times 10^{-9}.
\]
For these parameter values, the VEVs obtained from
minimization conditions are 
$v = 246.01$\,GeV, $w_1 = 1.0 \times 10^{-9}$\,GeV, 
$w_2 = 1.5 \times 10^{-9}$\,GeV. 

The Yukawa coupling matrices are fixed to be
\begin{eqnarray*}
h^{(1)}_{ij} &=&
\left(\begin{array}{ccc}
7.84 \times 10^{-4} & 4.55 \times 10^{- 3} & 8.71 \times 10^{- 5} 
\\ 
4.55 \times 10^{- 3}  & 1.19 \times 10^{- 2}  & 1.05 \times 10^{-2} 
\\ 
8.71 \times 10^{- 5}  & 1.05 \times 10^{- 2} & 2.22 \times 10^{- 2}
\end{array}\right),
\\
h^{(2)}_{ij} &=&
\left(\begin{array}{ccc}
2.0 \times 10^{-3} & 2.0 \times 10^{-3} & 2.0 \times 10^{-3} 
\\ 
2.0 \times 10^{-3}  & 1.0 \times 10^{-2}  & 1.0 \times 10^{-2} 
\\ 
2.0 \times 10^{-3}  & 1.0 \times 10^{-2} & 1.0 \times 10^{-2}
\end{array}\right).
\end{eqnarray*}
\subsection{Input parameters for Scenario 3: \\[-1mm]
$\mathrm{BR}(H_1^{++} \rightarrow W^+ W^+) \sim 
\mathrm{BR}(H_1^{++} \rightarrow \ell_i \ell_j)$}
\paragraph{BP 1:}

The input parameters for the scalar potential are
\[
a = -15625, \quad
\frac{1}{2}\,B = \left(
\begin{array}{cc}
58872.8 & -55110 \\ -55110 & 75002.3
\end{array} \right),
\]
\[
\frac{1}{4}\,D =  \left(
\begin{array}{cc}
1 & 1 \\ 1 & 1
\end{array} \right),
\;
\frac{1}{2}\,(E-H) = \left(
\begin{array}{cc} 0.8 & 0.95 \\ 0.95 & 1
\end{array} \right), 
\;
H = \left( \begin{array}{cc} 0.7 & 1 \\ 1 & 1 
\end{array} \right), 
\;
\frac{1}{4}\,F = \left( \begin{array}{cc} 0.7 & 0.5 \\ 0.5 & 1 
\end{array} \right)
\]
and 
\[
c = 0.2582, \; g = g' = 1, \; t_1 = -1 \times 10^{-5}, \; 
t_2 = -2 \times 10^{-5}.
\]
For these parameter values, the VEVs obtained from
minimization conditions are 
$v = 246.01$\,GeV, $w_1 = 0.59 \times 10^{-5}$\,GeV, 
$w_2 = 0.72 \times 10^{-5}$\,GeV. 

The Yukawa coupling matrices are fixed to be 
\begin{eqnarray*}
h^{(1)}_{ij} &=&
\left(\begin{array}{ccc}
5.15 \times 10^{-7} & 1.15 \times 10^{- 6} & 3.98 \times 10^{- 7} 
\\ 
1.15 \times 10^{- 6}  & 2.48 \times 10^{- 6}  & 2.22 \times 10^{- 6}  
\\ 
3.98 \times 10^{- 7}  & 2.22 \times 10^{- 6} & 4.21 \times 10^{- 6}
\end{array}\right),
\\
h^{(2)}_{ij} &=&
\left(\begin{array}{ccc}
1.0 \times 10^{-7} & 1.0 \times 10^{- 7} & 1.0 \times 10^{- 7}  
\\ 
1.0 \times 10^{- 7}  & 1.0 \times 10^{- 6}  & 1.0 \times 10^{- 6}  
\\ 1.0 \times 10^{- 7}  & 1.0 \times 10^{- 6} & 1.0 \times 10^{- 6}
\end{array}\right).
\end{eqnarray*}
\paragraph{BP 2:}

The input parameters for the scalar potential are
\[
a = -15627, \quad
\frac{1}{2}\,B = \left(
\begin{array}{cc}
58460.1 & -65200 \\ -65200 & 23292.4
\end{array} \right),
\]
\[
\frac{1}{4}\,D =  \left(
\begin{array}{cc}
1 & 0.9 \\ 0.9 & 1
\end{array} \right),
\;
\frac{1}{2}\,(E-H) = \left(
\begin{array}{cc} 0.8 & 1 \\ 1 & 1
\end{array} \right), 
\;
H = \left( \begin{array}{cc} 0.8 & 1 \\ 1 & 1 
\end{array} \right), 
\;
\frac{1}{4}\,F = \left( \begin{array}{cc} 0.8 & 0.5 \\ 0.5 & 1 
\end{array} \right)
\]
and 
\[
c = 0.2582, \; g = g' = 0.9, \; t_1 = -1 \times 10^{-4}, \; 
t_2 = -1.5 \times 10^{-4}.
\]
For these parameter values, the VEVs obtained from
minimization conditions are 
$v = 246.01$\,GeV, $w_1 = 1.0 \times 10^{-4}$\,GeV, 
$w_2 = 1.5 \times 10^{-4}$\,GeV. 

The Yukawa coupling matrices are fixed to be 
\begin{eqnarray*}
h^{(1)}_{ij} &=&
\left(\begin{array}{ccc}
7.84 \times 10^{-9} & 4.55 \times 10^{- 8} & 8.71 \times 10^{- 10} 
\\ 
4.55 \times 10^{- 8}  & 1.19 \times 10^{- 7}  & 1.05 \times 10^{-7} 
\\ 
8.71 \times 10^{- 10}  & 1.05 \times 10^{- 7} & 2.22 \times 10^{- 7}
\end{array}\right),
\\
h^{(2)}_{ij} &=&
\left(\begin{array}{ccc}
2.0 \times 10^{-8} & 2.0 \times 10^{-8} & 2.0 \times 10^{-8} 
\\ 
2.0 \times 10^{-8}  & 1.0 \times 10^{-7}  & 1.0 \times 10^{-7} 
\\ 
2.0 \times 10^{-8}  & 1.0 \times 10^{-7} & 1.0 \times 10^{-7}
\end{array}\right).
\end{eqnarray*}
\paragraph{BP 3:}

\[
a = -15627, \quad
\frac{1}{2}\,B = \left(
\begin{array}{cc}
77079.1 & -74990 \\ -74990 & 37283.5
\end{array} \right),
\]
\[
\frac{1}{4}\,D =  \left(
\begin{array}{cc}
1 & 0.9 \\ 0.9 & 1
\end{array} \right),
\;
\frac{1}{2}\,(E-H) = \left(
\begin{array}{cc} 0.82 & 0.9 \\ 0.9 & 0.82
\end{array} \right), 
\;
H = \left( \begin{array}{cc} 0.9 & 0.9 \\ 0.9 & 0.9 
\end{array} \right), 
\;
\frac{1}{4}\,F = \left( \begin{array}{cc} 0.9 & 0.45 \\ 0.45 & 0.9 
\end{array} \right)
\]
and 
\[
c = 0.2582, \; g = g' = 0.9, \; t_1 = -1 \times 10^{-4}, \; 
t_2 = -1.5 \times 10^{-4}.
\]
For these parameter values, the VEVs obtained from
minimization conditions are 
$v = 246.01$\,GeV, $w_1 = 1.0 \times 10^{-4}$\,GeV, 
$w_2 = 1.5 \times 10^{-4}$\,GeV. 

The Yukawa coupling matrices are fixed to be
\begin{eqnarray*}
h^{(1)}_{ij} &=& 
\left(\begin{array}{ccc}
7.84 \times 10^{-9} & 4.55 \times 10^{- 8} & 8.71 \times 10^{- 10} 
\\ 
4.55 \times 10^{- 8}  & 1.19 \times 10^{- 7}  & 1.05 \times 10^{- 7} 
\\ 
8.71 \times 10^{- 10}  & 1.05 \times 10^{- 7} & 2.22 \times 10^{- 7}
\end{array}\right),
\\
h^{(2)}_{ij} &=&
\left(\begin{array}{ccc}
2.0 \times 10^{-8} & 2.0 \times 10^{-8} & 2.0 \times 10^{-8} 
\\ 
2.0 \times 10^{-8}  & 1.0 \times 10^{-7}  & 1.0 \times 10^{-7} 
\\ 
2.0 \times 10^{-8}  & 1.0 \times 10^{-7} & 1.0 \times 10^{-7}
\end{array}\right).
\end{eqnarray*}
\paragraph{BP 4:}

The input parameters for the scalar potential are
\[
a = -15627, \quad
\frac{1}{2}\,B = \left(
\begin{array}{cc}
45594.7 & -55110 \\ -55110 & 17574.4
\end{array} \right),
\]
\[
\frac{1}{4}\,D =  \left(
\begin{array}{cc}
1 & 1 \\ 1 & 1
\end{array} \right),
\;
\frac{1}{2}\,(E-H) = \left(
\begin{array}{cc} 0.8 & 0.95 \\ 0.95 & 1
\end{array} \right), 
\;
H = \left( \begin{array}{cc} 0.7 & 1 \\ 1 & 1 
\end{array} \right), 
\;
\frac{1}{4}\,F = \left( \begin{array}{cc} 0.7 & 0.5 \\ 0.5 & 1 
\end{array} \right)
\]
and 
\[
c = 0.2582, \; g = g' = 1, \; t_1 = -1 \times 10^{-4}, \; 
t_2 = -1.5 \times 10^{-4}.
\]
For these parameter values, the VEVs obtained from
minimization conditions are 
$v = 246.01$\,GeV, $w_1 = 1.0 \times 10^{-4}$\,GeV, 
$w_2 = 1.5 \times 10^{-4}$\,GeV. 

The Yukawa coupling matrices are fixed to be 
\begin{eqnarray*}
h^{(1)}_{ij} &=&
\left(\begin{array}{ccc}
7.84 \times 10^{-9} & 4.55 \times 10^{- 8} & 8.71 \times 10^{- 10} 
\\ 
4.55 \times 10^{- 8}  & 1.19 \times 10^{- 7}  & 1.05 \times 10^{- 7} 
\\ 
8.71 \times 10^{- 10}  & 1.05 \times 10^{- 7} & 2.22 \times 10^{- 7}
\end{array}\right),
\\
h^{(2)}_{ij} &=&
\left(\begin{array}{ccc}
2.0 \times 10^{-8} & 2.0 \times 10^{-8} & 2.0 \times 10^{-8} 
\\ 
2.0 \times 10^{-8}  & 1.0 \times 10^{-7}  & 1.0 \times 10^{-7} 
\\ 
2.0 \times 10^{-8}  & 1.0 \times 10^{-7} & 1.0 \times 10^{-7}
\end{array}\right).
\end{eqnarray*}

\section{Expressions for doubly-charged scalar decay widths} 
\label{expressions}
In this part, we list the formulae for the decay rates of $H_1^{++}$ and
$H_2^{++}$. The masses of the doubly-charged scalars are denoted by
$M_{1,2}$ with $M_1 > M_2$ and those of the singly-charged scalars
by $\mu_{1,2}$ with $\mu_1 > \mu_2$. The mixing matrices $U$ and
$V$ are defined in equation~(\ref{UV}). With these quantities 
the decay rates for $H_1^{\pm\pm}$ and $H_2^{\pm\pm}$ can be evaluated as
\begin{align}
\Gamma(H_1^{++} \to \ell_i^+ \ell_j^+)
&=
\frac{1}{8\pi}
\left| h^{(1)}_{ij} U_{11} + h^{(2)}_{ij} U_{21} \right|^2 M_1 S_{ij},
\\[1mm]
\Gamma(H_1^{++} \to W^+ W^+)
&=
\frac{g^4 M_1^3}{16\pi m_W^4} \left| \left( U^\dagger w \right)_1 \right|^2
\left( \frac{3m_W^4}{M_1^4} - \frac{m_W^2}{M_1^2} + \frac{1}{4} \right) 
\beta\left( \frac{m_W^2}{M_1^2} \right), 
\\[1mm] 
\Gamma(H_1^{++} \to H_2^+ W^+ )
&=
\frac{g^2 M_1^3}{16\pi m_W^2} 
\left| \sum_{k=1,2} V_{k2}^* U_{k1} \right|^2
\left[
  \lambda\left( \frac{m_W^2}{M_1^2}, \frac{\mu_2^2}{M_1^2} \right)
\right]^{3/2}, 
\\[1mm]
\Gamma(H_2^{++} \to \ell_i^+ \ell_j^+)
&=
\frac{1}{8\pi}
\left| h^{(1)}_{ij} U_{12} + h^{(2)}_{ij} U_{22} \right|^2 M_2 S_{ij},
\\[1mm]
\Gamma(H_2^{++} \to W^+ W^+)
&=
\frac{g^4 M_2^3}{16\pi m_W^4} \left| \left( U^\dagger w \right)_2 \right|^2
\left( \frac{3m_W^4}{M_2^4} - \frac{m_W^2}{M_2^2} + \frac{1}{4} \right) 
\beta \left( \frac{m_W^2}{M_2^2} \right), 
\end{align} 
where 
\begin{equation}
S_{ij} = 
\left\{ \begin{array}{ccc}
1 & \mbox{for} & i \neq j, \\
1/2 & \mbox{for} & i = j.
\end{array} \right.
\end{equation}
The functions of $\lambda(x,y)$, $\beta(x)$ are 
defined as 
\begin{align}
\lambda(x,y) &= 1+x^2+y^2-2xy-2x-2y,\\
\beta(x)     &= \sqrt{\lambda(x,x)} = \sqrt{1-4x},
\end{align}
respectively.

\end{document}